\documentclass[prd,groupedaddress,twocolumn]{revtex4-1}
\usepackage{graphicx}
\usepackage{graphics}
\usepackage{amsmath}
\usepackage{amssymb}
\usepackage{mathtools} 

\begin{document}


\title{Classical and tree-level approaches to gravitational deflection in higher-derivative gravity }


\author{Antonio Accioly}

\email[]{accioly@cbpf.br}

\affiliation{Laborat\'{o}rio de F\'{\i}sica Experimental (LAFEX), Centro Brasileiro de Pesquisas F\'{i}sicas (CBPF), Rua Dr. Xavier Sigaud 150, Urca, 22290-180, Rio de Janeiro, RJ, Brazil}

\author{Jos\'{e} Helay\"{e}l-Neto}
\email[]{helayel@cbpf.br}

\affiliation{Laborat\'{o}rio de F\'{\i}sica Experimental (LAFEX), Centro Brasileiro de Pesquisas F\'{i}sicas (CBPF), Rua Dr. Xavier Sigaud 150, Urca, 22290-180, Rio de Janeiro, RJ, Brazil}

\author{Breno Giacchini}
\email[]{breno@cbpf.br}

\affiliation{Laborat\'{o}rio de F\'{\i}sica Experimental (LAFEX), Centro Brasileiro de Pesquisas F\'{i}sicas (CBPF), Rua Dr. Xavier Sigaud 150, Urca, 22290-180, Rio de Janeiro, RJ, Brazil}

\author{Wallace Herdy }
\email[]{wallacew@cbpf.br}

\affiliation{Laborat\'{o}rio de F\'{\i}sica Experimental (LAFEX), Centro Brasileiro de Pesquisas F\'{i}sicas (CBPF), Rua Dr. Xavier Sigaud 150, Urca, 22290-180, Rio de Janeiro, RJ, Brazil}



\date{\today}
\begin{abstract}
 Among  the so-called classical tests of general relativity (GR), light bending has been confirmed with an accuracy that increases as times goes by.  Here we study the gravitational deflection of photons within the framework  of classical and semiclassical  higher-derivative gravity (HDG) --- the only version of GR that is known up to now to be renormalizable along with its matter couplings.  Since our computations are restricted to scales much below the Planck cut-off we need not be afraid of the massive spin-2 ghost that haunts HDG. An upper bound on the constant related to the $R^2_{\mu \nu}$-sector of the theory is then found by analyzing --- from a classical and semiclassical viewpoint --- the deflection angle of a photon passing by the Sun. This  upper limit greatly improves that available in the literature. 
\end{abstract}



\maketitle

\section{Introduction}

General relativity (GR) is widely recognized as one of the keystones of Modern Physics. Notwithstanding,
it has not always been  adopted to set  bounds on physical parameters as it should. The pity of it is that the so-called classical tests of GR are moderately   used to estimate  limits on the constants that appear in relevant physical models.

 We remark that  among the aforementioned tests there is one, namely, light bending, which has been confirmed with great accuracy in the last two decades. This prediction of GR was first verified in 1919. Two separate expeditions to Sobral (Brazil) and Prince (Guinea), organized by Eddington and Dyson with  the aim of observing  the eclipse of May 29, 1919, reported deflections of $1.98 \pm 0.16''$ and $1.61 \pm 0.40''$, in reasonable accord with what Einstein thought would happen. Many measurements of the  gravitational deflection were then made in succeeding years, but the accuracy did not  really increase until the advent of   very long baseline  radio interferometry  in 1972, using quasar sources. In this vein, it is worth mentioning  two measurements of the solar gravitational deflection of radio waves made using  the aforementioned technique,  which are in excellent agreement with the prediction of GR. The first was made  by Lebach { \it  et al.}  \cite{1}; while  the other is owed to Fomalont, Kopeikin, Lanyi, and Benson \cite{2}. From the former   it was obtained a  deflection parameter $\gamma = 0.9996 \pm 0.0017 $, whereas  for the latter   
   $\gamma = 0.9998 \pm 0.0003$. Incidentally,  it is expected that a series of improved designed experiments with the Very Long Baseline Array  could increase the accuracy of the second measurement by at least a factor of 4 \cite{2}. 
	
	Interestingly enough, to the layman, light bending is one of the most impressive predictions made by Einstein. His celebrated formula $E=mc^2$ is in truth the only possible rival to the mentioned prediction  in popularity.  

On the other hand, higher-derivative  gravity models in ($3+1$) dimensions were suggested for the first time by Weyl \cite{3} and Eddington \cite{4}, being, roughly speaking, nothing but simple generalizations of GR obtained by enlarging Einstein Lagrangian via the scalars $R^2, R^2_{\mu \nu},$ and $R^2_{\mu \nu \alpha \beta}$. An interesting discussion about these classical systems can be found in the article by Havas \cite{5}. Later on it was shown that owing to the Gauss-Bonnet theorem only two of the terms above mentioned had to be added to Einstein Lagrangian.

However, only when it was proved that GR was nonrenormalizable within the standard perturbative scheme, did higher-derivative gravity  (HDG) --- up until then thought as a mere extension of Einstein's gravity  --- became indeed a prime candidate in the long and difficult search for a quantum gravity theory. In this vein, the seminal work done by Stelle in 1977 \cite{6} --- in which it was clearly shown that HDG is renormalizable along with its matter couplings --- is worthy of note. Unfortunately, this theory is nonunitary owing to the presence of a massive spin-2 ghost. 
By the way, in 1986, Antoniadis and Tomboulis \cite{7} claimed that the presence of a massive spin-2 ghost in the bare propagator is inconclusive, since this excitation is unstable. According  to them, the position of the complex poles in the dressed propagator is explicitly gauge dependent. Using standard arguments from quantum field theory they came to the conclusion that HDG theories are unitary.   Two years after Antoniadis and Tomboulis' article, Johnston \cite{8} proved that  the conjectures of these authors were not correct since the pair of  complex conjugate poles that appear in the resumed propagator are gauge independent, i.e., they are sedentary:  under a change in the gauge parameter they do not move. Therefore, HDG models are nonunitary.   

Before going on we shall discuss the common  misconception that  {\it singular}  higher-derivative models  can be discarded by appealing to the  Ostrogradski  theorem \cite{9}. For the sake of generality we consider higher-derivative systems in (N + 1) dimensions, with N=2, 3, ... . According to popular belief, Ostrogradski's result implies that there exists a linear instability in the Hamiltonian associated with all higher-derivative systems. This is a completely untrue assertion. Indeed, Ostrogradski only treated nonsingular models \cite{10}. Therefore, the only way of circumventing Ostrogradski's non-go theorem is by considering singular models, which is in accord with the conclusion reached by Woodard \cite{11}. An interesting example of this kind is the rigid relativistic particle studied by Plyushchay \cite{12}.

Now, since in this paper we are only interested in  higher-derivative gravity models, we remark that these systems are gauge invariant and, as consequence, are defined by singular Lagrangians \cite{10}. Thence, Ostrogradski's theorem does not apply to them, which does  not mean, of course,  that they are always ghost-free models. 

In (2+1) dimensions, for instance,  the BHT model (``new massive gravity''), which is defined by the Lagrangian

\begin{equation}
{\cal{L}}=\sqrt{g} \Bigg[- \frac{2R}{\kappa^2} + \frac{2}{\kappa^2 m^2_2} \Bigg(R^2_{\mu \nu} - \frac{3}{8} R^2 \Bigg) \Bigg]    ,\nonumber
\end{equation}

\noindent where $\kappa^2=4\kappa_3$, with $\kappa_3$ being   Einstein's constant in (2 +1) dimensions, and $m_2$($>0$) is a mass parameter,  has no ghosts at the tree level \cite{13,14,15,16}. Interestingly enough, $R + R^2$ gravity in $(N +1)$ dimensions, i.e., the model defined by the Lagrangian ${\cal{L}} = \sqrt{(-1)^{N-2}g}\Big[\frac{2R}{\kappa^2} + \frac{\alpha}{2}R^2 \Big]$, where $\kappa^2= 4\kappa_{N + 1}$, with   $\kappa_{N + 1}$  being Einstein's constant in $(N + 1)$ dimensions, and $\alpha$ is a free parameter, is also  tree-level unitary \cite{17}.

On the other hand, Sotiriou and Faraoni studied the so called $f(R)$ theories of gravity in (3 + 1)  dimensions  and at the classical level and came to the conclusion that ``theories of the form $f(R, R^2, R^2_{\mu \nu})$, contains, in general, a massive spin-2 ghost field in addition to the usual massless graviton and the massive scalar'' \cite {18}. Nevertheless, at the linear level,  these theories are stable \cite{19}. The reason why they do not explode is because the ghost cannot accelerate  owing to energy conservation. Another way of seeing this is by analyzing  the free wave solutions.  We remark that these models are not in disagreement with the result found by Sotiriou and Faraoni. Indeed, despite containing a massive spin-2 ghost, as asserted by these authors, the alluded  ghost cannot cause trouble.

Recently it was shown that at least in the  cases of specific  cosmological backgrounds, the unphysical  massive ghost  that haunts higher-derivative gravity in (3 + 1) dimensions and is present in the spectrum of this theory is not growing up as a physical excitation and remains in the vacuum state until the initial frequency of the perturbation is close to  the Planck scale. Accordingly, higher-derivative models of quantum gravity can be seen as very satisfactory effective theories of quantum gravity below the Planck cut-off \cite{20}.   

 Therefore, although HDG (higher-derivative gravity in (3 + 1) dimensions) is nonunitary in the framework of the usual quantum field theory, this does not imply that it must be rejected.

 We finish our digression by proving   that HDG systems  can be utilized at the tree level as effective field models at  scales far away from  the Planck scale.  Consider in this spirit the scattering at the tree level of a quantum particle  by HDG. Now, keeping in mind that the Lagrangian for HDG can be written as

\begin{equation}
{\cal{L}}_1= \sqrt{-g} \Big[ \frac{2}{\kappa^2}R + \frac{\alpha}{2}R^2 + \frac{\beta}{2}R^2_{\mu \nu} \Big],
\end{equation}

\noindent where $\kappa^2= 32 \pi G$, with $G$ being Newton's constant, and $\alpha$ and $\beta$ are free dimensionless coefficients, we promplyt find   that  the  associated propagator  is given in the de Donder gauge and in momentum space by \cite{21}

\begin{eqnarray}
D = && \Bigg[ \frac{1}{k^2} - \frac{1}{k^2 - m_2^2}\Bigg] P^{(2)} +\frac{2\lambda}{k^2}P^{(1)}  \nonumber \\ &&+   \frac{1}{2}\Bigg[    \frac{1}{k^2- m_0^2 }   -\frac{1}{k^2}  \Bigg ] P^{(0-s)} \nonumber \\  && + \Bigg[ \frac{4 \lambda}{k^2} + \frac{3m_0^2}{  2 k^2( k^2- m_0^2)} \Bigg] P^{(0-w)} \nonumber \\    &&+ \frac{\sqrt{3} m_0^2}{2  k^2 (k^2 - m_0^2)} \Bigg[ P^{(0-sw)} + P^{(0-ws)} \Bigg],
\end{eqnarray}

\noindent where $\lambda$ is a gauge parameter, $\{P^{(1)}, P^{(2)}, ...\; , P^{(0-ws)} \}$ is the set of  the usual Barnes-Rivers operators, and 

\begin{eqnarray}
m_2^2 \equiv  -\frac{4 }{\beta  \kappa^2}, \;\;\; m_0^2 \equiv \frac{2}{\kappa^2 [3\alpha +  \beta]}. 
\end{eqnarray}

\noindent We are assuming, of course, that $m^2_2 >0 \; (\beta <0)$ and $m^2_0 >0 \; (3 \alpha + \beta >0)$, so as to avoid tachyons in the model.

Let us them show that HDG is tree-level unitary at the aforementioned scales. To accomplish this, we make use of a method pioneered by Veltman \cite{22} that has been extensively used since it was conceived.
Veltman's prescription consists in saturating the propagator with conserved external currents  and computing afterward the residues at the simple poles of the alluded saturated   propagator ($SP$). If the residues at all the  poles are positive or null, the model is tree-level unitary, but if at least one of the residues is negative, the system is nonunitary  at the tree level.

 The saturated propagator in momentum space is in turn given by

\begin{eqnarray} 
SP(k)&&= T_{\mu \nu}(k)D^{\mu \nu, \alpha \beta}(k)T_{\alpha \beta}(k) \nonumber \\   &&= \frac{A}{k^2} - \frac{B}{k^2 - m^2_2} + \frac{C}{k^2 - m^2_0}.  \nonumber
\end{eqnarray}

\noindent Here $$A\equiv T^2_{\mu \nu} - \frac{T^2}{2}, \; B \equiv T^2_{\mu \nu} - \frac{T^2}{3}, \; C \equiv \frac{T^2}{6}.$$

Let us then suppose that $k^2 \ll m^2_2$. Consequently, $$ SP(k)= \frac{A}{k^2} + \frac{C}{k^2- m^2_0} + {\cal{O}}\Big( \frac{k^2}{m^2_2} \Big).$$  Now, bearing in mind that for a massless graviton $$\Big(T^2_{\mu \nu} - \frac{T^2}{2}\Big)\Big \vert_{k^2=0} >0 \; \; {\mathrm{(See \;Ref. \;17 )}},$$ we come to the conclusion that $$Res(SP)|_{k^2=0} >0, \; Res(SP)|_{k^2 = m^2_0} >0.$$ Therefore, at the scale at  hand, HDG is unitary at the tree level and, as a consequence,  the massive spin-2 ghost is  completely harmless.

Now, owing to the great interest  this gravity theory  has aroused in the literature, it should be important  to analyze the issue of the gravitational deflection in its framework and, using this result, to  find bounds on its  free constants. This is precisely our goal in this paper. To do that we shall study the gravitational deflection of a photon passing by the Sun in the context of the gravity theory at hand using classical and  tree-level approaches. Since the $R^2$-sector of the model does not contribute  anything to the  gravitational deflection, we cannot  estimate an upper bound on the constant concerning this sector of the system by analyzing the light bending; nevertheless, we shall discuss  in  the latter section of the paper, in  passing, how to find a bound on this constant by using another classical test of GR.  On the other hand, by suitable combining  the classical and semiclassical results concerning solar gravitational deflection,  we will be able   to estimate an upper limit on the constant of the  $R^2_{\mu \nu}$-sector.  The latter greatly improves the current  bound  available in the literature.

The article is organized as follows. In Sec. II we study the gravitational deflection of light  by the Sun using a classical approach, while in Sec. III we analyze the solar   gravitational bending of a photon at the  tree level.  An upper bound on the constant of the $R^2_{\mu \nu}$-sector of the theory is then obtained in Sec. IV by judiciously joining together the classical and tree-level results. Our conclusions are presented in Sec. V.

We use natural units throughout and our Minkowski metric is diag(1, -1, -1, -1).

\section{ Light bending in classical higher-derivative gravity }

To begin with, we solve  the linearized field equations related to HDG.

The field equations concerning the  Lagrangian density

\begin{equation}
{\cal{L}}= {\cal{L}}_1 - {\cal{L}}_{\mathrm{M}},
\end{equation}

\noindent where ${\cal{L}}_{\mathrm{M}}$ is the Lagrangian density for matter, are

$$\frac{2}{\kappa^2}G_{\mu \nu} + \frac{\beta }{2}\Big[ -\frac{1}{2}g_{\mu \nu}R^2_{\rho \lambda} + \nabla_\mu \nabla_\nu R+ 2R_{\mu \rho \lambda \nu}R^{\rho \lambda}$$  $$- \frac{1}{2}g_{\mu \nu}\Box R - \Box R_{\mu \nu} \Big] + \frac{\alpha}{2}\Big[ -\frac{1}{2}g_{\mu \nu} R^2  + 2R R_{\mu \nu}$$ $$+ 2 \nabla_\mu \nabla_\nu R - 2 g_{\mu \nu} \Box R \Big]  + \frac{1}{2} \Theta_{\mu \nu} = 0, $$

\noindent where $\Theta_{\mu \nu}$ is the energy-momentum tensor.

From the above equation  we promptly obtain  its linear approximation doing exactly    as in Einstein's theory. We write

  \begin{equation}
	g_{\mu \nu}= \eta_{\mu \nu} + \kappa h_{\mu \nu},
	\end{equation}
	\noindent  and then linearize the equation at hand via (5), which results  in the following

$$\Big( 1- \frac{\beta \kappa^2}{4}\Box \Big) \Big[ -\frac{1}{2} \Box h_{\mu \nu}  + \frac{1}{6 \kappa} R^{(\mathrm{lin})} \eta_{\mu \nu} \Big] + \frac{1}{2}(\Gamma_{\mu, \nu}$$ $$+ \Gamma_{\nu, \mu})= \frac{\kappa}{4} \Big( T_{\mu \nu} - \frac{1}{3}T \eta_{\mu \nu} \Big),$$

\noindent where $$ R^{(\mathrm{lin})}= \frac{\kappa}{2}\Box h - \kappa {\gamma^{\mu \nu}}_{, \mu \nu},$$ $$  \gamma_{ \mu \nu} \equiv h_{\mu \nu} - \frac{1}{2}\eta_{\mu \nu} h,$$
$$\Gamma_\mu \equiv \Big(1 - \frac{\beta \kappa^2}{4} \Box \Big) {\gamma_{\mu \nu}}^{ ,\nu} - \Big( \alpha + \frac{\beta}{2} \Big)\frac{\kappa}{2} {R^{(\mathrm{lin})}}_{, \mu}.$$

\noindent  Note that indices are raised (lowered) using $\eta^{\mu \nu}$ ($\eta_{\mu \nu}$). Here $T_{\mu \nu}$ is the energy-momentum tensor of special relativity.

It can be shown that it is always possible to choose  a coordinate system such that the gauge conditions, $
\Gamma_\mu=0$, on the linearized metric hold. Assuming that these conditions are satisfied, it is straightforward to show that the general solution of the linearized  field equations is given by \cite{23,24}

\begin{equation}
h_{\mu \nu}= h^{({\mathrm{E}})}_{\mu \nu}- \phi \eta_{\mu \nu} + \psi_{\mu \nu},
\end{equation}

\noindent where $h^{({\mathrm{E}})}_{\mu \nu}$ is the solution of  linearized Einstein's equations in the de Donder gauge, i.e., 

 \begin{equation}
\Box h^{({\mathrm{E}})}_{\mu \nu}= \frac{\kappa}{2} \Big[ \frac{T \eta_{\mu \nu}}{2} - T_{\mu \nu} \Big], \;\;\; \gamma^{{({\mathrm{E}}) \;,\nu}}_{\mu \nu} =0,  \nonumber
\end{equation}
 
$$\gamma^{{({\mathrm{E}})}}_{\mu \nu} \equiv h^{({\mathrm{E}})}_{\mu \nu} - \frac{1}{2}\eta_{\mu \nu} h^{({\mathrm {E}})},$$

\noindent while $\phi$ and $\psi_{\mu \nu}$ satisfy, respectively, the equations

$$\Big(\Box + m^2_0  \Big) \phi =\frac{\kappa T}{12},$$ $$\Big(
\Box + m^2_2 \Big) \psi_{\mu \nu}=\frac{\kappa}{2} \Big[T_{\mu \nu} - \frac{1}{3}T \eta_{\mu \nu }\Big], \; \Box \psi= {\psi_{\mu \nu}}^{, \mu \nu}.$$

\noindent It is worth noting that in  this very special gauge the equations for $\psi_{\mu \nu}, \; \phi, \;$ and $h^{({\mathrm{E}})}_{\mu \nu} $ are totally decoupled. As a result, the general solution to the
linearized  field equations reduces to an algebraic sum of the solutions of the  equations  concerning the three mentioned fields.

Solving the latter for a pointlike particle of mass $M$ located at ${\bf r=0}$ and having, as a consequence, an energy momentum tensor $T_{\mu \nu}= M\eta_{\mu 0} \eta_{\nu 0}\delta^3({\bf r})$, we find

\begin{equation}
h_{\mu \nu}(r)= h^{(\mathrm{E})}_{\mu \nu}(r) + h^{(\mathrm{R^2})}_{\mu \nu}(r) + h^{\mathrm{(R_{\mu \nu}^2)}}_{\mu \nu}(r) ,
\end{equation}

\noindent with

$$h^{(\mathrm{E})}_{\mu \nu}(r)= \frac{M\kappa}{16 \pi} \Big[ \frac{\eta_{\mu \nu}}{r} - \frac{2\eta_{\mu 0} \eta_{\nu 0}}{r}\Big],$$ 
 
$$h^{(\mathrm{R^2})}_{\mu \nu}(r)= \frac{M\kappa}{16 \pi} \Big[ -\frac{1}{3}\frac{e^{-m_0 r}}{r} \eta_{\mu \nu} \Big],$$ 

$$h^{\mathrm{(R_{\mu \nu}^2)}}_{\mu \nu}(r)=\frac{M\kappa}{16 \pi} \Big[ - \frac{2}{3}\frac{ e^{-m_2 r }}{r}\eta_{\mu \nu} + 2 \frac{e^{- m_2 r}}{ r}\eta_{\mu 0} \eta_{\nu 0}\Big].$$

\noindent Note that for $m_0, \; m_2 \rightarrow \infty$, the above solution reproduces the solution of linearized Einstein field equations in the de Donder gauge, as it should. We also remark that employing  a method recently developed, that relies on Feynman path integral and  allows 
the computation of the $(N + 1)$-dimensional interparticle potential energy in a
straightforward way \cite{21,25}, we can  trivially obtain  the potential energy for the interaction of two masses $M_1, \; M_2$ separated for a distance $r$. Utilizing this prescription, we find 

$$E(r)= M_1 M_2 G \Big[ -\frac{1}{r} - \frac{1}{3} \frac{e^{-m_0 r}}{r} + \frac{4}{3}\frac{e^{-m_2 r}}{r}    \Big],$$

\noindent which agrees asymptotically with Newton's potential energy, as expected.

We are now ready to discuss the light bending owed to the gravitational field  sourced by the mass $M$. Suppose, in this spirit, a photon with momentum $p_\mu$ coming from infinity with an impact parameter $b$ (See Fig. 1). The net change in $p_\mu $ while it passes through the aforementioned gravitational field is given by

\begin{equation}
\Delta p_\mu = \frac{\kappa}{2}p^\alpha \int_{- \infty}^{\infty}{\partial_{\mu} h_{\alpha \beta} dx^\beta},
\end{equation}

\noindent where the integration is performed along the approximately straight line trajectory of the photon. As a consequence, the displacement along the approximately straight ray and the momentum are, respectively,

$$dx^\mu\approx(dx^1, dx^1, 0, 0), \;\; p^\mu \approx (p^1, p^1, 0, 0).$$

\begin{figure}[h!]
	\centering
		\includegraphics[scale=0.5]{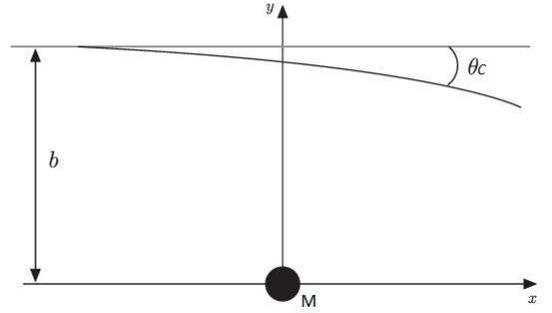}
	\caption{
	Geometry of the light bending.}
	\label{fig}
\end{figure}
Inserting these quantities into Eq. (8), we obtain

\begin{eqnarray}
\Delta p_2= \frac{\kappa}{2}p^1 \int_{-\infty}^{\infty}{\partial_y[h_{00} + h_{11}]dx^1},
\end{eqnarray}

\noindent which can be written as

\begin{eqnarray}
\Delta p_2= \frac{\kappa}{2} p^1 \int_{-\infty}^{\infty}{\Big[\Big( \frac{d}{dr}(h_{00} + h_{11}) \Big) \frac{\partial r}{\partial y} \Big]\Big \vert_{y=b} dx}.
\end{eqnarray}

\noindent With the result (7), we can rewrite Eq. (10) simply as

\begin{eqnarray}
\Delta p_2 = && \frac{M \kappa^2 b}{16 \pi} p^1 \int_{-\infty}^{\infty}\Bigg[ \frac{1}{(x^2 + b^2)^{\frac{3}{2}}}  \nonumber \\ &&- \frac{1 + m_2(x^2 +b^2)^{\frac{1}{2}}}{(x^2  + b^2)^{\frac{3}{2}}} e^{-m_2(x^2 + b^2)^{\frac{1}{2}}}\Bigg] dx.
\end{eqnarray}

Therefore, the classical deflection angle, i.e., $\theta_{\mathrm{C}} \equiv  \Big \vert \frac{\Delta p_y}{p_x} \Big \vert=  \Big \vert \frac{ -\Delta p_2}{p^1} \Big \vert $, can be computed through the expression

\begin{eqnarray}
\theta_{\mathrm{C}}  = \theta_{\mathrm{E}}  -  \frac{M \kappa^ 2 b}{16 \pi}\int_{-\infty}^{\infty}\frac{1 + m_2(x^2 +b^2)^{\frac{1}{2}}}{(x^2  + b^2)^{\frac{3}{2}}} e^{-m_2(x^2 + b^2)^{\frac{1}{2}}}dx, \nonumber
\end{eqnarray}

\noindent where  $\theta_{\mathrm{E}}$ is Einstein's deflection angle.

At this point, some comments are in order.

\begin{enumerate}

\item It is trivial to see from the above result that $\theta_{\mathrm{C}} \rightarrow \theta_{\mathrm{E}}$ as $m_2  \rightarrow \infty$ ($|\beta| \rightarrow 0$). In other words, in this limit we recover Einstein's prediction for the light bending. That is the reason why the integration constant related to the mentioned equation is zero. In addition, the limit for $m_2 \rightarrow 0$ ($|\beta| \rightarrow \infty$) clearly shows  the absence of deflection. Both results are physically consistent. 
 
\item The scalar excitation of mass $m_0$ does not contribute at all to the light bending. Why is this so? Because the metric concerning linearized $R +  R^2$ gravity --- the theory obtained by linearizing the field equations related to the Lagrangian ${\cal L}= \sqrt{-g}[ \frac{2}{\kappa^2}R + \frac{\alpha}{2}R^2 - {\cal L}_{\mathrm{M}}]$ --- is conformally related to  linearized GR. Indeed, denoting the solution to the linearized $R + R^2$ gravity by $g^{(R + R^2)}_{\mu \nu}$,  we promptly obtain $ g^{(R + R^2)}_{\mu \nu} \equiv \eta_{\mu \nu } + \kappa g^{(R + R^2)}_{\mu \nu}= (1 - \kappa \phi)g^{(E)}_{\mu \nu}, $ where, of course, terms of order $\kappa^2 $ were neglected.

\item A quick glance at the equation at hand  shows that the dependence of $\theta_{\mathrm{C}}$ on $|\beta|$ is dominated by the exponential term, which suggests that the transition from the Einsteinian limit to the no-deflection scenario might be localized in a well-defined interval. Outside this domain, $\theta_{\mathrm{C}}$ is practically constant. Thence, we come to the conclusion that $0 \leq \theta_{\mathrm{C}} \leq \theta_{\mathrm{E}}$. 

\end{enumerate}

Numerical integration allows the evaluation of the deflection angle for different values of $|\beta|$.  The result for a light ray just grazing the Sun is depicted in Fig. 2. We point out that the transition interval occurs for $ 10^{84} \lesssim|\beta | \lesssim 10^{88}$. Therefore, in order not to conflict with the prediction of GR for the  solar gravitational deflection  which, incidentally,  has been exhaustively  tested experimentally  with great success, $|\beta| < 10^{84}$.

\begin{figure}[h!]
	\centering
		\includegraphics[scale=0.8]{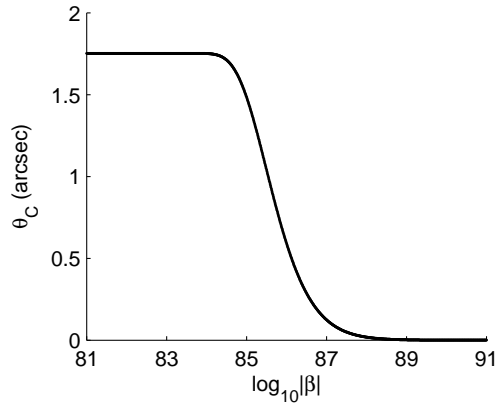}
	\caption{Deflection angle $\theta_{\mathrm{C}}$ as a function of ${\mathrm{log}}_{10}|\beta|$ for light rays just grazing the Sun in classical HDG.}
	\label{fig}
\end{figure}

\section{Gravitational deflection in tree-level hdg}
Semiclassical gravity is based on the following type of approximation scheme: the metric is considered as  a classical field, predetermined by the gravitational field equations which in our case  are  those of HDG; besides, the energy content of some particles and/or fields are often neglected. In addition, the spacetime, which is nothing but a fixed background, is determined, uniquely, for example, by a huge, static, point mass $M$. Incidentally, the mass $M$ is huge in comparison to the energy of the other particles and/or fields that either exert a tiny influence on the space time or do not affect it at all. And more,  the classical gravitational field  interacts with particles that are quantum in nature. As is well known, the results found via a semiclassical gravity theory are more comprehensive than those obtained from the corresponding classical one. In fact, at the classical level we deal with  structureless particles, while at the tree level  we are involved with  quantum particles. Of course, in the classical limit the former results reduce to the latter. As far as GR  is concerned, interesting examples related to this subject can be found, for example,  in    \cite{26,27,28,29,30}.

Let us then analyze the gravitational deflection of a photon within the context of tree-level HDG. Consider, in this vein, the scattering of  this photon  by  the external gravity field  (7). The Feynman amplitude for this process is given by (See Fig. 3)

\begin{figure}[here!]
	\centering
		\includegraphics[scale=0.4]{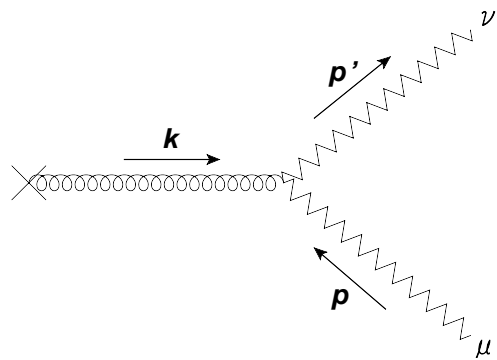}
	\caption{ Photon scattering by an  external gravitational field. Here $|{\bf{p}}|= |{\bf{p'}}|$.}
	\label{one-loop: fig}
\end{figure}

\begin{eqnarray}
{\cal{M}}_{r r'} =&& \frac{1}{2} \kappa h^{\lambda \rho}_{\mathrm{ext}}({\bf{k}})\Bigg[-\eta_{\mu \nu} \eta_{\lambda \rho}pp' + \eta_{\lambda \rho}p'_\mu p_\nu + 2\Big(\eta_{\mu \nu}p_\lambda p'_\rho  \nonumber \\  &&- \eta_{\nu \rho}p_\lambda p'_\mu  \nonumber - \eta_{\mu \lambda }p_\nu p'_\rho + \eta_{\mu \lambda} \eta_{\nu \rho}pp'\Big) \Bigg]\epsilon^\mu_r({\bf{p}}) \epsilon^\nu_{r'}({\bf{p'}}) \nonumber,
\end{eqnarray}

\noindent where $\epsilon^\mu_r({\bf{p}})$ and $\epsilon^\nu_{r'}({\bf{p'}})$ are  the polarization vectors for the initial and final photons, respectively, which satisfy the completeness relation

\begin{equation}
\sum_{r=1}^{2}\epsilon^\mu_r({\bf{p}})\epsilon^\nu_{r}({\bf{p}})= -\eta^{\mu \nu} - \frac{p^\mu p^\nu}{(pn)^2} + \frac{p^\mu n^\nu + p^\nu n^\mu}{pn},
\end{equation}

\noindent where $n^2=1$. Here $h^{\lambda \rho}_{\mathrm{ext}}({\bf{k}})$ is the momentum space gravitational field, namely,

\begin{equation}
h^{\lambda \rho}_{\mathrm{ext}}({\bf{k}})= \int{d^3{\bf{r}} e^{-i{\bf{k}}\cdot {\bf{r}}}h^{\lambda \rho}_{\mathrm{ext}}({\bf{r}})}.
\end{equation}

\noindent Thence,

\begin{eqnarray}
h^{\lambda \rho}_{\mathrm{ext}}({\bf{k}})= h^{{\mathrm{(E)}} \lambda \rho}_{\mathrm{ext}}({\bf{k}}) + h^{ {\mathrm{(R^2_{\mu \nu})}}\lambda \rho}_{\mathrm{ext}}({\bf{k}}) + h^{{\mathrm{(R^2)}}\lambda \rho}_{\mathrm{ext}}({\bf{k}}),
\end{eqnarray}

\noindent with

$$h^{{\mathrm{(E)}} \mu \nu}_{\mathrm{ext}}({\bf{k}}) = \frac{\kappa M}{4{\bf{k}}^2}\eta^{\mu \nu} - \frac{\kappa M}{2} \frac{ \eta^{\mu 0} \eta^{\nu 0}}{\bf{\bf{k}}^2},$$

$$h^{{\mathrm{(R^2_{\alpha \beta})}} \mu \nu}_{\mathrm{ext}}({\bf{k}})= - \frac{\kappa M}{6} \frac{\eta^{\mu \nu}}{{\bf{k}}^2 + m^2_2} +\frac{\kappa M}{2}\frac{\eta^{\mu 0} \eta ^{\nu 0}}{{\bf{k}}^2 + m^2_2}, $$

$$h^{{\mathrm{(R^2)}} \mu \nu}_{\mathrm{ext}}({\bf{k}}) = -\frac{\kappa M}{12} \frac{\eta^{\mu \nu}}{{\bf{k}}^2 + m^2_0}. $$

The unpolarized cross-section can then be written as

\begin{eqnarray}
\frac{d\sigma}{d \Omega}&&= \frac{1}{(4 \pi)^2} \frac{1}{2}\sum_{r} \sum_{r'}{\cal{M}}^2_{r r'} \nonumber \\ &&= \frac{1}{(4 \pi)^2 }\frac{\kappa^4 M^2 E^4(1 + \cos{\theta})^2}{16} \Bigg[-\frac{1}{{\bf{k}}^2} + \frac{1}{{\bf{k}}^2 + m^2_2} \Bigg]^2, \nonumber
\end{eqnarray}

\noindent where $E$ is the energy of the incident photon and $\theta $ is the scattering angle. 

For small angles the preceding equations reduces to

\begin{eqnarray}
\frac{d\sigma}{d \Omega}= 16 G^2 M^2 \Big[-\frac{1}{\theta^2} + \frac{1}{\theta^2 + \frac{m^2_2}{E^2}} \Big]^2.
\end{eqnarray}

\noindent This result signals an energy-dependent scattering. 

 It is easy to see from (15) that 

$$ \frac{d \sigma}{d \Omega} \rightarrow 0, \;\; {\mathrm{if}} \; \frac{ m_2}{E} \rightarrow 0,$$

$$\frac{d \sigma}{d \Omega} \rightarrow  \Big(\frac{4GM}{\theta^2} \Big)^2,\;\; {\mathrm{if}} \; \frac{ m_2}{E} \rightarrow \infty;$$

\noindent in other words, if $ {\frac{m_2}{E}} \rightarrow 0$, there is no scattering, whereas   if  $
\frac{m_2}{E} \rightarrow  \infty$, we recover Einstein's  standard cross-section, as expected.

Now, in order to arrive at a classical  particle trajectory from (15), we compare the classical and the tree-level cross-section formulas \cite{31,32}

\begin{eqnarray}
\frac{d\sigma}{d \Omega}= 16 G^2 M^2 \Bigg[-\frac{1}{\theta^2} + \frac{E^2}{E^2 \theta^2 + m^2_2} \Bigg]^2= -\frac{rdr}{\theta d\theta}.
\end{eqnarray}

Performing the integration we promptly find that for a photon just grazing the Sun the above equation gives the following result 

\begin{eqnarray}
\frac{1}{\theta^2_{\mathrm{E}}}= \frac{1}{\theta^2} + \frac{1}{\lambda^2 + \theta^2} +  \frac{2}{\lambda^2} \ln{ \frac{\theta^2}{\lambda^2 + \theta^2}},
\end{eqnarray}

\noindent with $\lambda^2 \equiv \frac{m^2_2}{E^2}$. We call attention to the fact that the integration constant related to the this equation was temporarily omitted for the sake of a cautious and meticulous analysis of the behavior of the $\theta$-dependent function we shall perform in the following; of course,  it will  be restored in due course. To do the aforementioned investigation in a consistent way, we define beforehand  a function $\gamma = \gamma \Big(\frac{m^2_2 }{E^2} \Big) >0 $ so that the $\theta $ angle can be written as 

\begin{eqnarray}
\theta = \gamma \frac{m_2}{E}.
\end{eqnarray}

As a result, Eq. (17) can be rewritten as

\begin{eqnarray}
\frac{\lambda^2}{\theta^2_{\mathrm{E}}}= \frac{1}{\gamma^2} + \frac{1}{1 + \gamma^2} + 2 \ln{\Big( \frac{\gamma^2}{1 + \gamma^2} \Big)} \equiv f(\gamma),
\end{eqnarray}

\noindent or

\begin{eqnarray}
\frac{\theta^2}{\theta^2_{\mathrm{E}}}= \gamma^2 f(\gamma).
\end{eqnarray}

We remark that since $f$ is a monotonically decreasing function of $\gamma$ having as image the interval $(0, +\infty)$, it can be shown that is always possible to find a solution to (17) in the form (18). In addition, $\gamma$ is a decreasing function, implying that the limit $\lambda \rightarrow 0$ corresponds, for instance,  to let $\gamma \rightarrow \infty$ in Eq. (20).

We are now ready to analyze the behavior of $\theta$ at different situations. It is straightforward to see that for  a fixed energy $E$, $\theta \rightarrow \theta_{\mathrm{E}}$ as $|\beta| \rightarrow 0$ and $\theta \rightarrow 0$ as $|\beta| \rightarrow \infty$. The former regime recovers Einstein's one, as desirable, while the latter shows that for a sufficient large $|\beta|$ no deflection occurs. We also point  out that $0 \leq \theta \leq \theta_{\mathrm{E}}$ since $\gamma^2 f(\gamma) \leq 1$.

The repulsive contribution to the bending, which arrives from the $R^2_{\mu \nu}$-sector, is energy dependent as it is evident from (17). Inasmuch as $|\beta|$ is thought to be a (universal) constant, it is worthwhile to analyze the behavior of the scattering angle  for a fixed $|\beta|$ and different values of $E$. It is obvious that in this scenario $\theta \rightarrow \theta_{\mathrm{E}}$ in the low energy (classical) limit, and $\theta \rightarrow 0$ for sufficiently energetic photons, suggesting that the more energetic a photon is, the less it will deviate. Let us then show that this is indeed the case by  finding the solutions to Eq. (17) for visible light. In Fig. 4 it is shown how $\theta$ behaves for different values of $|\beta|$. A quick glance at this graphic allows us to conclude that for a fixed $E$, the scattering angle for visible light is approximately constant for almost  all values of $|\beta|$, making a transition from $\theta \approx \theta_{\mathrm{E}}$ to $\theta \approx 0$ in a well defined interval of width $\Delta|\beta| \approx 10^{10}$.

\begin{figure}[here!]
	\centering
		\includegraphics[scale=0.54]{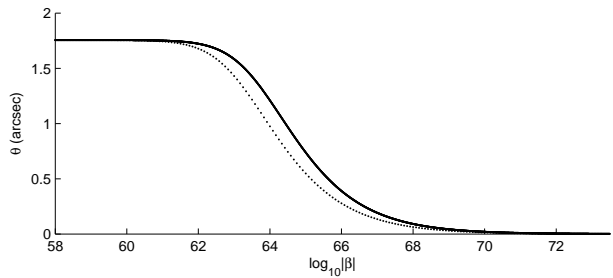}
	\caption{$\theta_{\mathrm{red}}$ (continuum  line) and  $\theta_{\mathrm{violet}}$ (dotted line) for photons passing by the Sun as a function of $\log_{10}{|\beta |}$  in semiclassical HDG.}
	\label{fig}
\end{figure}

Accordingly, in the framework of tree-level HDG the visible spectrum, whose wavelength ranges from 4000 to 7000 (\AA), would spread over an angle $|\Delta \theta|$, where $|\Delta \theta| \equiv |\theta_{\mathrm{violet}} - \theta_{\mathrm{red}}|$. Let us then evaluate $|\Delta \theta| $ for different values of $|\beta|$ using Eq. (17).  The results are shown in Fig. (5). 

\begin{figure}[here!]
	\centering
		\includegraphics[scale=0.538]{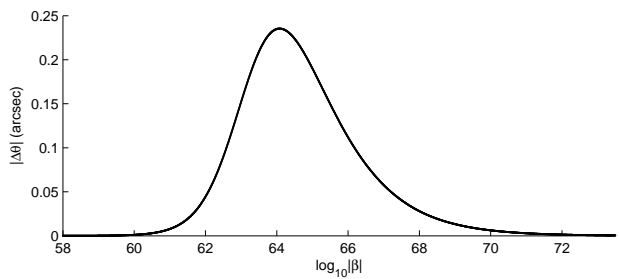}
	\caption{$|\Delta \theta|$  as a function  of $\log_{10}{|\beta|}$ for photons passing by the Sun's limb  in the context  of  tree-level HDG.}
	\label{fig} 
\end{figure}

A cursory inspection of this  graph allows us to conclude that for  $ 61  \lesssim \log_{10}{|\beta|} \lesssim 71$  the spread of the visible spectrum would in principle be observable. Actually, we ought to  expect a tiny value for  $|\Delta \theta |$ at the Sun's limb in order not to conflict with well established results of  experimental general relativity. Consequently, if $|\beta|< 10^{61}  $,   the visible spectrum spread  would be practically imperceptible and the deviation angle would be very close to the Einstein one.  Accordingly, we come to the conclusion that in order to agree with the currently measured values for visible light, $|\beta|< 10^{61}  $. We point out that this bound was estimated by noting that the  gravitational rainbow predicted by tree-level HDG is incompatible with today measurements.  Of course, the mentioned limit would be modified if we have made use of photons with wavelength outside the domain of the visible light. 

\section{Smooth transition from the semiclassical context to the classical one}

The most striking difference between the classical and semiclassical approaches is, perhaps,  the fact that the repulsive interaction owed to  the $R^2_{\mu \nu}$-sector  depends on the photon energy that interacts with the gravitational field. Since in the classical realm the gravitational field acts on structureless particles, gravity scatters light of all wavelengths in the same way; nonetheless, in the tree-level scenario more energetic photons are more repelled and, as consequence, less deflected.   

Now, a point that deserves a careful attention is the subtle divergence between these scenarios  at low energy: the classical limit of the semiclassical  theory does not match that of classical HDG. In fact, in the classical model,  whatever the energy of the light ray is, no scattering will occur if $|\beta | > 10^{89}$. On the other hand, the analysis at the tree-level does not impose  any upper bound at all on the  interval of the $|\beta|$ transition; consequently, it is always possible to find  a small $E$  so that $\theta$ is arbitrarily close  to $\theta_{\mathrm{E}}$, even if $|\beta|> 10^{89}$. A way out of this difficulty, would be to add a non trivial integration constant to Eq. (17) which, as a result, assumes the  form 

\begin{eqnarray}
\frac{1}{\theta^2_{\mathrm{E}}}= \frac{1}{\theta^2} + \frac{1}{\lambda^2 + \theta^2} +  \frac{2}{\lambda^2} \ln{ \frac{\theta^2}{\lambda^2 + \theta^2}} - \Omega,
\end{eqnarray}
 
\noindent Indeed, choosing  $\Omega$ as a function only  of $|\beta|$, it is possible to make it to give
a negligible contribution in the range of energies   such that  the transition occurs for $|\beta| \lesssim 10^{85}$, and to be relevant for the photons which  make its transition above this interval.

Let us then  compare the deflection angles computed in both frameworks, i.e., $\theta$ and $\theta_{\mathrm{C}}$, requiring  furthermore  that $\theta \rightarrow \theta_{\mathrm{C}}$ if $E \rightarrow 0$. Using the limit calculated in Sec. 3, Eq. (21) reduces to

\begin{equation} 
\frac{1}{\theta^2_{\mathrm{E}}}= \frac{1}{\theta^2} - \Omega,
\end{equation}

\noindent whose solution is

\begin{equation}
\theta = \theta_{\mathrm{E}} \Big(1 + \Omega \theta^2_{\mathrm{E}} \Big)^{-\frac{1}{2}}.
\end{equation}

\noindent Now, imposing  that $\theta = \theta_{C}$, we promptly obtain

\begin{equation}
\Omega= \frac{1}{\theta^2_{\mathrm{C}}} - \frac{1}{\theta^2_{\mathrm{E}}}.
\end{equation}

Our next step is to check whether the Einsteinian limit ($|\beta| \rightarrow 0$) is indeed consistent. In  Sec. 2 we got that $\theta_{\mathrm{C}} \rightarrow \theta_{\mathrm{E}}$ if $|\beta| \rightarrow 0$, and, as a result, $\Omega \rightarrow 0$. Moreover, it is easy to see that $\Omega \gg \theta^{-2}_{\mathrm{E}}$ if ${|\beta| \gtrsim 10^{88}}$. Therefore,  the limit $|\beta| \rightarrow 0$ remains unchanged, and Einstein solar deflection angle is recovered, as it should. We point out that for $|\beta| < 10^{85}$
the  integration constant $\Omega$ can be simply neglected. For larger values of $|\beta|$,  nevertheless, $\Omega$ increases too quickly forcing $\theta \rightarrow 0$ even for low energy photons. Besides, the classical results are recovered in the classical limit. In Figs. 6 and 7 we display some values of  $\Omega$  for different $|\beta|$s.

\begin{figure}[here!]
	\centering
		\includegraphics[scale=0.7]{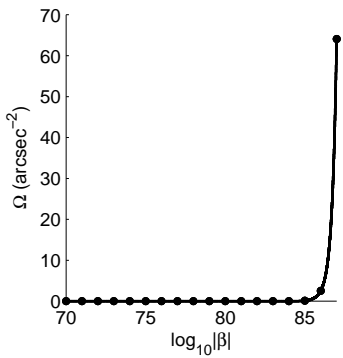}
	\caption{$\Omega$ as a function of $|\beta|$. It is worth noting that $\Omega$ is roughly zero for $|\beta|< 10^{85}$, where it starts the approximately double exponential growth depicted in Fig. 7.}
	\label{one-loop: fig}
\end{figure}

Two comments fit here.

\begin{enumerate}

\item We have shown  in Sec. 3 that for visible light the transition from $\theta_{\mathrm{E}}$ to 0 in the absence of the integration constant took place for $|\beta| \in (10^{61}, 10^{71})$. Making use of these values an upper bound on $|\beta|$ was estimated. We remark that  this result remains unchanged since within the mentioned domain, as we have proved, $\Omega $ can be taken to be equal zero.

\item In order to allow the deflection angle computed at the tree level to agree   in the classical limit with   the result found directly via the classical approach, we have to appeal to the integration constant $\Omega$ (See  (22)).  On the other hand, for $|\beta| < 10^{85}$  this constant is tiny, implying that it can be left out of any computation if we take  the current experimental accuracy into account. Now, since  $|\beta| < 10^{84} $ is the upper bound on $|\beta |$ found classically, and $|\beta| < 10^{61}$ is that arising from the tree-level computations, we come to the conclusion that the constant $\Omega$
can be simply neglected.
\end{enumerate}

\begin{figure}[here!]
	\centering
		\includegraphics[scale=0.7]{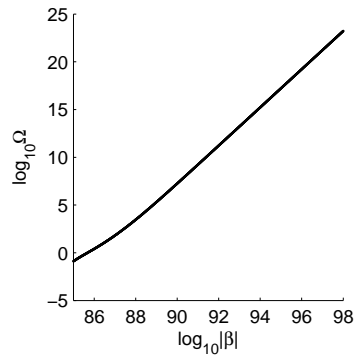}
	\caption{$\Omega$ (in arcs$^{-2}$) for some values of  $|\beta|$. Here the $\Omega$-axis scale is also logarithmic in order  to show  its quick increase for $|\beta| > 10^{85}$.
	}
	\label{one-loop: fig}
\end{figure}

\section{Final remarks}
We have shown that the photon propagation in the framework of  tree-level HDG is dispersive. From the analysis of the energy-dependent contribution coming from the photons passing by the Sun, it was possible to estimate an upper bound on $|\beta|$, namely,  $|\beta| < 10^{61}$. Let us then compare this upper limit with that available in the literature. Using the interesting measures of Long \cite{33}, Stelle \cite{34} found that $m_2 \approx 1 \times 10^{-4} cm^{-1}$. From this value Donoghue \cite{35} estimated that $|\beta| \leq 10^{74}$. Therefore, our bound lowered the accepted limit on $|\beta|$ by thirteen orders of magnitude.

We call attention to the fact that the measurements made in the radio band, despite its precision and accuracy,  do not improve the limit on $|\beta|$ we have found.
In fact, since less energetic photons undergo a greater bending, the transition interval from $\theta = \theta_{\mathrm{E}}$ to $\theta = 0$ occurs for the measured  radio waves about 10 orders of magnitude above the visible waves. However, if gravitational deflection  measurements in the X-ray or ultraviolet bands were available, we could certainly improve the limit on $|\beta|$. Unfortunately, it is a very hard task to separate the signs  present in these wavelengths from those emitted by  the Sun. Accordingly, we come to the conclusion that the bound we have obtained is the best limit one can found using the gravitational  deflection measurements available nowadays.

To conclude, we mention that we have  estimated  a bound on the constant of the  $R^2$-sector of HDG using the  accurate  experimental results  we have at our disposal  today concerning the gravitational
 red shift of the spectral lines. This limit will be  published elsewhere  \cite{36}. Interestingly enough, the cited classical text of GR was the first  conceived    by Einstein to verify his theory but the last  to   have reliable experimental results.

\begin{acknowledgments}
A. A. is very grateful to I. Shapiro and S. Dias for fruitful discussions. The authors acknowledge financial  support from CNPq and  FAPERJ.
\end{acknowledgments}

\end{document}